\documentclass[lettersize,journal]{IEEEtran}
\usepackage{amsmath,amsfonts}
\usepackage{algorithmic}
\usepackage{algorithm}
\usepackage{array}
\usepackage[caption=false,font=normalsize,labelfont=sf,textfont=sf]{subfig}
\usepackage{textcomp}
\usepackage{stfloats}
\usepackage{url}
\usepackage{verbatim}
\usepackage{graphicx}
\usepackage{cite}
\usepackage{booktabs}
\usepackage{xcolor}
\usepackage{tcolorbox}

\usepackage{xcolor}
\usepackage{multirow}

\newcommand{\lqy}[1]{{#1}}

\newcommand{\qyminor}[1]{{#1}}

\usepackage{wrapfig}

\newcounter{finding}
\usepackage{mdframed}
\newcommand{\finding}[1]{\refstepcounter{finding}
  \vspace{0.5mm}
 \begin{mdframed}[linecolor=gray,roundcorner=12pt,backgroundcolor=gray!15,linewidth=3pt,innerleftmargin=2pt, leftmargin=0cm,rightmargin=0cm,topline=false,bottomline=false,rightline = false]
  \textbf{Answer to RQ\arabic{finding}:} #1
 \end{mdframed}
 \vspace{0.5mm}
}

\begin{document}

\title{Condor: A Code Discriminator Integrating General Semantics with Code Details}

\author{Qingyuan Liang, Zhao Zhang, Chen Liu, Zeyu Sun*, Wenjie Zhang, Yizhou Chen, Zixiao Zhao, Qi Luo, Wentao Wang, Yanjie Jiang, Yingfei Xiong, Lu Zhang*
        % <-this % stops a space
\thanks{Qingyuan Liang, Zhao Zhang, Chen Liu, Yizhou Chen, Zixiao Zhao,  Wentao Wang, Yanjie Jiang, Yingfei Xiong, and Lu Zhang are with
Key Lab of HCST (PKU), MOE; SCS, Peking University,
Beijing 100871, China (e-mail: liangqy@pku.edu.cn; zhangzhao2019@pku.edu.cn;
cissieliu@stu.pku.edu.cn; yizhouchen@stu.pku.edu.cn; 2301111987@stu.pku.edu.cn; wwt@stu.pku.edu.cn; yanjiejiang@pku.edu.cn; xiongyf@pku.edu.cn; zhanglucs@pku.edu.cn).}% <-this % stops a space
\thanks{Zeyu Sun is with the National Key Laboratory of Space Integrated Information System, Institute of Software, Chinese Academy of Sciences,
Beijing, China (e-mail: zeyu.zys@gmail.com).}
\thanks{Wenjie Zhang is with the National University of Singapore,
Singapore, (e-mail: wjzhang@nus.edu.sg).}
\thanks{Qi Luo is with the Southern University of Science and Technology,
Shenzhen, China (e-mail: 12232440@mail.sustech.edu.cn).}
\thanks{*Corresponding authors: Zeyu Sun and Lu Zhang}
}



\maketitle

\begin{abstract}
LLMs demonstrate significant potential across various software engineering tasks. However, they still face challenges in generating correct code on the first attempt when addressing complex requirements. Introducing a discriminator to select reliable outputs from multiple generated results is an effective way to enhance their reliability and stability.
Currently, these discriminators fall into two categories: execution-based discriminators and non-execution-based discriminators. Execution-based discriminators face flexibility challenges due to difficulties in obtaining test cases and security concerns, while non-execution-based discriminators, although more flexible, struggle to capture subtle differences in code details.
To maintain flexibility while improving the model’s ability to capture fine-grained code details, this paper proposes Condor.
We first design contrastive learning to optimize the code representations of the base model, enabling it to reflect differences in code details. Then, we leverage intermediate data from the code modification process to further enrich the discriminator’s training data, enhancing its ability to discern code details.
Experimental results indicate that on the subtle code difference dataset~(i.e., CodeNanoFix), Condor significantly outperforms other discriminators in discriminative performance: Condor~(1.3B) improves the discriminative F1 score of DeepSeek-Coder~(1.3B) from 67\% to 73\%. In discriminating LLM-generated outputs, Condor~(1.3B) and Condor~(110M) raise the Pass@1 score of Llama-3.1-Instruct~(70B) on the CodeNanoFix dataset from 52.64\% to 62.63\% and 59.64\%, respectively. 
\lqy{Moreover, Condor demonstrates strong generalization capabilities on the APPS, MBPP, and LiveCodeBench datasets. }
For example, Condor~(1.3B) improves the Pass@1 of Llama-3.1-Instruct~(70B) on the APPS dataset by 147.05\%.
\end{abstract}


\section{Introduction}\label{sec:intro}
\IEEEPARstart{W}{ith} the widespread application of large language models~(LLMs), their potential for tasks in software engineering~(e.g., code generation and code repair) receives significant attention~\cite{guo2024deepseekcoder, dubey2024llama3, team2024codegemma, llm4understanding, chatgpt2024, scis_deeplearning}. However, they typically struggle to produce correct code on the first attempt when addressing complex requirements such as advanced algorithmic problems~\cite{apps, evaluating_code_quality, khan2023assessingpromisepitfallschatgpt, ji2023surveyhallucination}.
Existing approaches typically employ an automated discriminator to select a correct result from multiple outputs generated by LLMs, improving the reliability and stability of LLMs when handling complex problems~\cite{coderanker}.

Currently, these approaches for code discrimination fall into two main categories: execution-based approaches and non-execution-based approaches. Execution-based approaches require running the code before making a discriminative decision, utilizing feedback from execution as additional detailed information to enhance the discriminator's performance. However, execution-based models face challenges in flexibility. 
First, executing code poses challenges because (1) many real-world downstream reasoning tasks lack test cases~\cite{softwaretesting}, (2) LLM-generated code may contain incomplete snippets~\cite{chen2021evaluatinglargelanguagemodels}, and (3) the testing environments and other dependencies are not always accessible~\cite{coderanker, copilot}. 
Second, even when execution is possible, running LLM-generated code raises safety concerns.
On the other hand, non-execution-based approaches do not execute code and rely solely on the details presented in the code text, providing greater flexibility and applicability across different application scenarios. 

\begin{figure*}[t]
  \centering
  % \vspace{-0.13cm}
  % \setlength{\abovecaptionskip}{10pt}
  \includegraphics[scale=1]{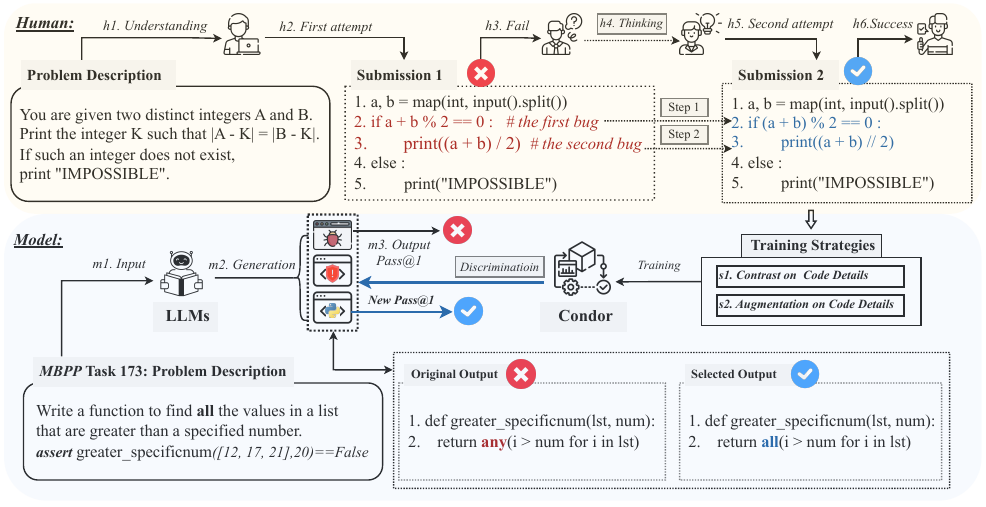}
  % \vspace{-0.2cm}
  \caption{An example illustrates how humans write correct code through thinking, and how models rely on a discriminator to select the correct code. The upper section illustrates the interaction with the code evaluation system, where the user attempts to submit the code twice. The lower section displays the correct code selection by a discriminator, where the model may not be able to generate the correct answer on its first attempt. Thus, it is necessary to employ a discriminator to enhance the reliability of the generated outputs.}
  \label{example}
  \vspace{-0.5cm}
\end{figure*}

However, existing non-execution-based discriminators typically leverage general code representation capabilities without adapting to the nuanced details of the programming process~\cite{coderanker,coder_reviewer_reranking}. 
Challenges remain in 
% quantifying a discriminator's ability to detect fine-grained code differences and 
improving the discriminator's effectiveness at capturing intricate code details. 
In the development scenario, the functionality of code often depends on key minor changes, such as variable updates, conditional logic modifications, or adjustments in function calls. 
For example, as shown in Figure~\ref{example}, the upper section illustrates a problem description about finding a number K equally distant from two distinct numbers A and B. Submission 1 displays buggy code, while submission 2 shows the correct code after the user's modification. Notably, only three characters differ between the two submissions~(i.e., `(', `)', and `/'), yet the functionality of the first submission remains incorrect.
% 模型也会犯同样的错误
Similarly, outputs generated by LLM can also contain errors caused by subtle differences~\cite{xu2024subtle_error, shi2024code_error}, demonstrating that such mistakes are not unique to humans. For example, as shown below in Figure~\ref{example}, the model~(i.e., Llama-3.1-Instruct-70B) generates two outputs for Task 173 in the MBPP~\cite{mbpp} benchmark, one of which is incorrect due to a minor substitution, where `all' is mistakenly replaced with `any'.
Existing discriminators primarily rely on embeddings provided by other general-purpose code models as a basis for representation. This limits their ability to capture subtle code details, consequently reducing the effectiveness of identifying similar sample pairs in LLM-generated outputs.

To keep the flexibility and enhance the detail-capturing capabilities of the code discriminator, we propose Condor, a non-execution-based discriminator that integrates general semantics with code details. To address the challenges that non-execution-based approaches face in capturing subtle code differences, we design an embedding-level contrastive learning strategy~\cite{contrastive_learning} and a data-level intermediate data mining strategy for Condor.
In our embedding-level contrastive learning strategy, we first sample code pairs with high textual similarity for the same problem. Then, we design two loss functions during the training process to progressively bring the embeddings of correct code samples closer together, while increasing the embedding distance between correct and incorrect code snippets.
Condor refines the base model's code embeddings through contrastive learning to better distinguish between correct and incorrect code representations.
For the data-level intermediate data mining strategy, we utilize data from multiple submissions for the same problem to identify intermediate data not covered by the existing training set. 
This strategy aims to capture intermediate data that the evaluation system may not have recorded during the user’s problem-solving process.
For example, in the upper section of Figure~\ref{example}, after the user's thinking process~(i.e., h4. Thinking), the user submits a new code snippet~(i.e., submission 2) with modifications in two steps. 
After the modification in Step 1, the code still contains a bug. This partially modified version, which has not yet undergone the Step 2 modification, is referred to as partially fixed intermediate data.
In real-world scenarios, users may submit a new code version only after careful consideration, and the partially fixed code for previous submissions may not always be explicitly recorded in the evaluation system.
By leveraging these partially fixed buggy code snippets, Condor can incorporate the incremental details in the programmer's thinking process to further enhance its discriminative capabilities in specific tasks.
As shown in the lower section of Figure~\ref{example}, by integrating these two strategies for capturing code details, Condor aims to automatically select the correct code from multiple outputs when the LLM struggles to generate the correct code on the first attempt.

\lqy{
We implement two Condor versions~(i.e., 110M and 1.3B) to balance performance and practicality: the smaller model enables low-latency ranking in resource-constrained settings~\cite{codebert2020,coderanker,codet52021}, while the larger model offers stronger accuracy, providing a better trade-off than single-shot inference with much larger models~\cite{guo2024deepseekcoder, grammarcoder}.
}
When training a discriminator, we find that current datasets are insufficient for fully quantifying the discriminator's ability to discern differences in code details~\cite{humaneval, commitpack_muennighoff2023octopack}. Thus, we construct the CodeNanoFix dataset, which includes a series of samples consisting of an algorithmic problem description, pairs of code with subtle errors, and the corresponding correct version.
First, we apply the Condor to the classification task, evaluating its ability to recognize subtle code differences.
The results show that general code generation models struggle to accurately distinguish these fine-grained differences, often resulting in imbalances in precision or recall, leading to suboptimal overall performance. In contrast, Condor excels across precision, recall, and F1-score, significantly outperforming other models. For example, Condor~(110M) improves the F1-score from around 67\% to 72.34\%. 
Additionally, Condor~(1.3B) demonstrates substantial improvements in discriminative capability when selecting the correct code from multiple generated outputs, achieving an increase in Pass@1 accuracy of Llama-3.1-Instruct~(70B) from 52.64\% to 62.63\%.
Second, to validate Condor’s generalization ability, we test it on the APPS~\cite{apps}, the MBPP~\cite{mbpp}, and the LiveCodeBench~\cite{jain2024livecodebench} datasets. The results indicate that Condor consistently maintains strong performance across different datasets. 
\lqy{
For example, on the APPS dataset, Condor~(1.3B) raises the Pass@1 of Llama-3.1 Instruct~(70B) from 10.16\% to 25.10\%, on MBPP, Condor~(1.3B) improves Pass@1 from 71.40\% to 75.20\%, and on LiveCodeBench, Condor~(1.3B) improves Pass@1 from 28.97\% to 30.90\%.
}
Finally, through ablation studies, we demonstrate that the two strategies for capturing code details in Condor are crucial to its overall performance, further validating their effectiveness in detecting subtle code differences.

We summarize our contributions as follows:
\begin{itemize}

\item \qyminor{We propose Condor~\cite{condor_github}, a novel, non-execution-based code discriminator that incorporates embedding-level and data-level code details to discriminate and filter code generated by LLMs.}
\item We construct CodeNanoFix, a new code dataset focusing on bugs caused by subtle code changes, to evaluate whether the discriminators can distinguish fine-grained differences in code.
\item We conduct extensive experiments, and results show that Condor effectively differentiates between code that appears similar in form but differs in functionality.
\end{itemize}

\section{Condor}
% In this section, we first explain how to train Condor to enhance the discriminator's ability to capture code details. Then, we introduce how to evaluate the discriminator's effectiveness.

\subsection{Overview of Condor}
% CL: 翻前sample，翻后loss
% CLS：翻前data，翻后prob

\begin{figure*}[t]
  \centering
  % \vspace{-0.13cm}
  % \setlength{\abovecaptionskip}{10pt}
  \includegraphics[width=\linewidth]{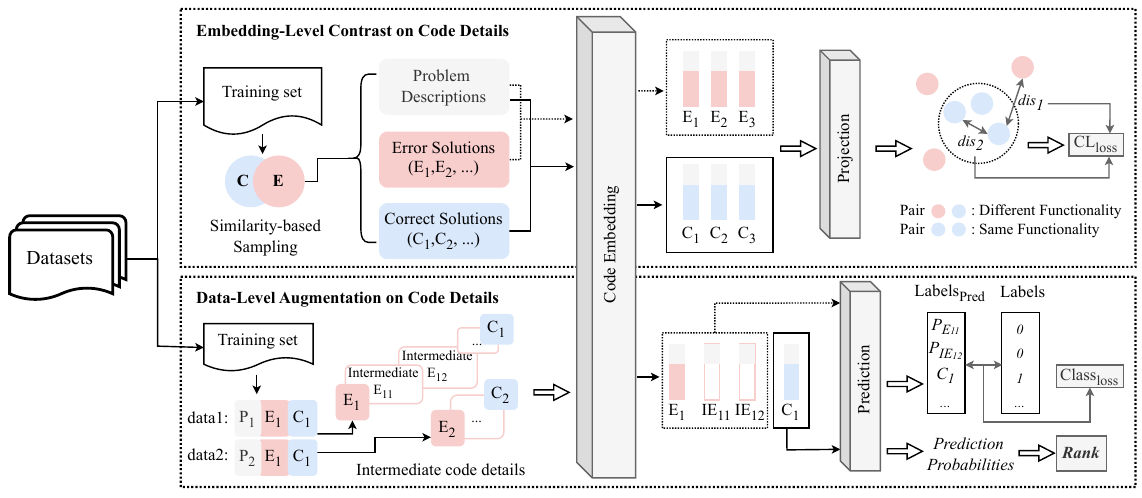}
  \caption{The Condor overview consists of two main components: contrastive learning at the embedding level to capture code details (upper section), and data-level augmentation through intermediate code, which supplements code details that are not recorded in existing datasets (lower section). The `C' denotes the correct code that passes all test cases, while `E' indicates the error code that fails some test cases.}
  \label{overview}
  \vspace{-0.5cm}
\end{figure*}

Condor is a model designed to discriminate the correctness of code purely based on its text, without the need to execute the code to gather additional information. This gives Condor greater flexibility in real-world applications, especially when test cases are unavailable. 

% We train two versions of Condor with different parameter scales~(i.e., Condor-110M and Condor-1.3B) based on these strategies. We choose to use CodeT5's encoder and DeepSeek-Coder-1.3B as the base models to train Condor at different parameter scales for two main reasons. First, both models demonstrate strong performance in various software engineering tasks and are widely used in the field, providing a solid foundation. Second, compared to larger models like DeepSeekCoder-Instruct-6.7B or Code-Llama-3.1-Instruct-70B, they are relatively small, which helps us control the parameter size of the discriminator, making training and deployment more efficient across a variety of tasks.

However, distinguishing between code samples that are very similar in text becomes more challenging for the discriminator, as it may struggle to fully utilize code details to differentiate between code quality or functionality. 
To address this challenge, Condor, as illustrated in Figure~\ref{overview}, employs two strategies to enhance its ability to capture fine-grained code details.
\begin{itemize}
    \item Embedding-Level Contrast on Code Details: We design a contrastive learning strategy to capture code details at the embedding level.
    \item Data-Level Augmentation on Code Details: We utilize intermediate code details, which may be missing from existing datasets, to augment the training dataset.
\end{itemize}
% Below, we present these strategies in detail.

Our approach, named Condor (literally meaning an eagle) not only searches for prey in the sky~(metaphorically representing the general semantics of code) but also dives from the sky to capture prey on the ground~(metaphorically representing the details of code). This hunting activity metaphorically illustrates how Condor~(\textbf{Co}de and its \textbf{N}uanced Integrated \textbf{D}iscriminat\textbf{or}) can focus on both the overall structure and the fine-grained differences in code, thereby enhancing the performance of the discriminator.

\subsection{Embedding-Level Contrast on Code Details}
% CL
We employ a contrastive learning~\cite{contrastive_1, contrastive_2, zhao2024spotting, chen2024improving} strategy to make the model aware that textually similar yet functionally different code snippets should have distinct representations at the embedding level. 
% 对比学习在干什么，有的reviewer不懂
% Contrastive learning is typically used to enhance the similarity between embeddings of samples within the same category while reducing the similarity between embeddings of samples from different categories.
Contrastive learning, known for enhancing the similarity between embeddings within the same category while reducing the similarity between those from different categories, is well-suited to our task.
To fully utilize this mechanism, we design a similarity-based data sampling strategy tailored to code discrimination. 
By incorporating this design, we leverage contrastive learning to effectively capture subtle code differences, transforming them into meaningful parameter updates to refine the model's code embedding representations.
% To leverage this mechanism for capturing subtle differences in code details, we design a similarity-based data sampling approach that focuses on buggy code snippets. By applying contrastive learning to these samples, the model translates these detailed code differences into meaningful parameter updates, refining its code embedding representations.
% To effectively capture the differences in code details, we design a similarity-based data sampling approach targeting code snippets containing bugs. Building on this, we use contrastive learning to transform these detailed code differences into meaningful parameter updates for the model, optimizing its code embedding representations.

\subsubsection{Similarity-based Sampling}
In contrastive learning, we focus on comparing samples with a high degree of textual similarity in the data sampling process, enabling the discriminator to differentiate between textually similar but functionally distinct code snippets.
Each data sample in contrastive learning is represented as \(\langle p, c_{a}, c_{b}, label_{a,b} \rangle\), where \(p\) represent the problem description, \(c_{a}\) and \(c_{b}\) are two textually similar code snippets, and \(label_{a,b}\) indicates the relationship between them. 
Each contrastive learning sample takes into account both the problem description and the code representation, ensuring that the code embeddings are compared within the context of a specific problem.
Based on whether the code can pass test cases~(C: correct code that passes all test cases, E: error code that fails some test cases), the relationship between \(c_{a}\) and \(c_{b}\) can be classified into four categories: C-C, C-E, E-E, and E-C.
% 解释E-C and C-E
Due to the symmetry between E-C and C-E in classification, we treat them as a single category.
Since we primarily focus on the subtle differences between correct and incorrect code, we emphasize comparisons between C-C and E-C pairs. 
This helps the model recognize which code details affect functionality and which do not. Thus, the labels \(label_{a,b}\) are divided into two key types: C-C~(i.e., pairs with the same functionality) and E-C (pairs with different functionality). 
We assign a label of 1 to C-C pairs and 0 to E-C pairs and use these labeled data to train the supervised model for contrastive learning.

\subsubsection{Contrastive Learning}
Contrastive learning aims to make the embeddings of the same functionality pairs closer and those of different functionality pairs farther apart. 
This helps the model differentiate between correct code~(C) and error code~(E) at the representation level. 

To achieve this goal, we first map the code and its associated problem representation into points in a two-dimensional space. We then quantify the similarity between code snippets by calculating the Euclidean distance between these points. 
We design a loss function based on Euclidean distance, aiming to maximize the distance between functionally correct and incorrect code pairs while minimizing the distance between functionally correct code pairs.
The contrastive learning loss function is defined as follows:
% \[
% \mathcal{L} = \frac{1}{N} \sum_{i=1}^{N} \left[ label_{a,b} \cdot d(x_{a}^{i}, x_{b}^{i})^2 + (1 - label_{a,b}) \cdot \max(0, m - d(x_{a}^{i}, x_{b}^{i}))^2 \right],
% \]
\begin{multline*}
\mathcal{L} = \frac{1}{N} \sum_{i=1}^{N} \bigg[ label_{a,b} \cdot d(x_{a}^{i}, x_{b}^{i})^2 \\
+ (1 - label_{a,b}) \cdot \max(0, m - d(x_{a}^{i}, x_{b}^{i}))^2 \bigg],
\end{multline*}

where \( N \) is the total number of training samples, \(label_{a,b} \) is a binary label for each pair, and \(label_{a,b} = 1 \) indicates the same functionality pairs (C-C) and \( label_{a,b} = 0 \) indicates different functionality pairs (E-C). \( d(x_{a}^{i}, x_{b}^{i}) \) is the Euclidean distance between the two embeddings \( x_{a}^{i} \) and \( x_{b}^{i} \). \( m \) is the margin that ensures dissimilar pairs are separated by at least a certain distance.

This loss encourages the same functionality pairs to have smaller distances while different functionality pairs are pushed apart by at least the margin \( m \).
For example, in Figure~\ref{overview}, this loss function encourages the model to increase the distance \( dis_1 \) (between the embeddings of correct and buggy code pairs) and decrease the distance \( dis_2 \) (between the embeddings of correct pairs). This optimization helps the model learn to distinguish between functionally correct and incorrect code, even when the text is similar.

\subsection{Data-Level Augmentation on Code Details}
% Intermediate
In real-world development, programmers often make multiple changes while attempting to fix a bug, but not all intermediate code versions are recorded. For example, as shown in Figure~\ref{example}, the user submitted two versions of the code—an incorrect version~(on the left) and a corrected version~(on the right). It can be observed that transitioning from the incorrect to the correct code requires two-step modifications, and simply fixing one part would not have resulted in a proper solution. 
For example, a potentially unrecorded version that combines the first two lines of the correct code~(Submission 2) with the last three lines of the error code~(Submission 1) would still be incorrect.

\subsubsection{Intermediate Code Details}
To ensure the model to recognize that the code must be correct in all details, we explore the potential paths from incorrect to correct code by incorporating partially fixed intermediate versions. 
We leverage these intermediate bug-fix data~(e.g., as shown in the Figure~\ref{example}, step 1 is modified in line 2 while step 2 remains unchanged in line 3) samples to enhance the training set. By introducing these partially fixed versions, the model gains a deeper understanding of the nuanced differences in code modifications, improving its ability to perceive and discriminate code correctness. 
This strategy effectively simulates the real bug-fixing process, helping the model know whether the current modifications are sufficient to correct the code.

To describe this process more clearly, we use the following notation to represent the various stages of modification:
Let \( E_1 \) represent the initial incorrect code and \( C_1 \) represent the final correct code. We first compute \( \text{diff}(E_1, C_1) \), which represents all the modifications required to transform the incorrect code \( E_1 \) into the correct code \( C_1 \).
Next, we decompose \( \text{diff}(E_1, C_1) \) into \( n \) independent modification steps:
\[
\text{diff}(E_1, C_1) = \{ \Delta_1, \Delta_2, \dots, \Delta_n \},
\]
Each \( \Delta_i \) represents a part of the modifications needed to transform \( E_1 \) into \( C_1 \). 
% \Delta对应到实际代码中是什么？
Specifically, each \( \Delta_i \) is represented as a diff hunk, which may consist of multiple consecutive lines of changes. If the difference between the two code versions involves only one hunk, the lines within that hunk are further split, treating each line as an individual modification.
We then apply these modifications sequentially to \( E_1 \), generating a series of intermediate code versions \( IE_{11}, IE_{12}, \dots, IE_{1n} \):
\[
IE_{11} = E_1 + \Delta_1
\]
\[
IE_{12} = IE_{11} + \Delta_2
\]
\[
\vdots
\]
\[
C_1 = IE_{1(n-1)} + \Delta_n
\]
Through this process, we generate multiple intermediate versions of the code as it progresses from incorrect to correct. Each intermediate version \( IE_{1i} \) represents the code state after a partial modification. 
After generating all the intermediate code versions \( IE_{11}, IE_{12}, \dots, IE_{1n} \), we further perform deduplication against the original training data to ensure no duplicate code snippets are introduced. Once deduplication is complete, these filtered intermediate code versions are added to the original training set, creating an augmented dataset. 
This approach allows the model to learn the subtle changes from partial to full fixes, improving its ability to understand and handle code details.
This approach generates more training data with finer granularity and provides the model with multi-level information about the code repair process, helping the model better capture the progressive transformations from incorrect to correct code.

\subsubsection{Discriminating}
As a code discriminator, Condor intends to select functionally correct code from a set of candidate code snippets. Based on the embedding representations enriched with code details through contrastive learning, and the augmented training dataset containing intermediate code details, we train a binary discriminative model to determine whether a given code is functionally correct. Let the candidate code set be \( \{ C_1, C_2, \dots, C_k \} \).

During training, we first define the label \( y_i \in \{0, 1\} \), where \( y_i = 1 \) indicates that the code \( C_i \) is functionally correct, and \( y_i = 0 \) indicates that the code is functionally incorrect.  Then, we construct pairs \( \langle C_i, y_i \rangle \) to train a binary classification model as the discriminator. The cross-entropy loss function is applied to optimize the model, improving its ability to distinguish between correct and incorrect code. The loss function is defined as:
\[
\mathcal{L} = - \frac{1}{k} \sum_{i=1}^{k} \left( y_i \log(p_i) + (1 - y_i) \log(1 - p_i) \right),
\]
% where \( p_i = \sigma(f(\mathbf{V}_i)) \) represents the likelihood that code \( C_i \) is functionally correct, \( f(\mathbf{V}_i) \) is the logits value output by the model, and \( \sigma(\cdot) \) is the sigmoid activation function.
where \( p_i = \sigma(f(\mathbf{V}_i)) \) represents the likelihood that code \( C_i \) is functionally correct, \( f(\mathbf{V}_i) \) is the logits value output by the model, and \( \sigma(\cdot) \) is the sigmoid activation function. Specifically, \( p_i \) is computed by applying the sigmoid function to the embedding logits, which maps the output to a probability range between 0 and 1, indicating the confidence of the model in the functional correctness of \( C_i \).

During inference, given a set of candidate code snippets \( \{ C_1, C_2, \dots, C_k \} \), the model outputs the probability \( p_i \) that each code snippet \( C_i \) is functionally correct. We then select the candidate code with the highest probability as the most likely correct code:
\[
C_{\text{selected}} = \arg\max_{i} p_i
\]

\subsection{CodeNanoFix Dataset}
% To effectively train a discriminator integrating code detail, it is essential to use code snippets that are textually similar but functionally distinct to test their ability to distinguish subtle differences. 
To better train a discriminator capable of detecting subtle code differences, it is essential to use data samples from real-world programming scenarios where programmers have written textually similar but functionally distinct code. 
% A key metric for evaluating discriminators is their ability to differentiate between code snippets that are textually similar yet functionally different~(i.e., producing different outputs on the same test cases).
% why not existing dataset
Existing code generation datasets~\cite{humaneval,mbpp,apps} are less suitable because these datasets lack real-world fine-grained modifications, and their training set relies on code generated by specific LLMs, making the discriminator's effectiveness dependent on the particular LLM used.
Additionally, identical functionality can be implemented in various ways, resulting in generated code that is not always textually similar.
Therefore, we construct a new repair-related dataset, CodeNanoFix, comprising training, validation, and test sets to support the training of discriminators capable of detecting subtle changes in code.
% what dataset
% Unlike code generation, which relies on generating code based on natural language prompts, 
\subsubsection{Data Collection}
CodeNanoFix consists of tuples of problem description, buggy code, and correct code, designed to evaluate human-written code with subtle differences.
For example, Figure~\ref{example} illustrates two code snippets that are nearly identical in text but differ in functionality: `Submission 1' on the left is incorrect~(not passing the test cases), while `Submission 2' on the right is a correct solution corresponding to the problem description~(passing the test cases). 
% how to construct
We collect the data samples from CodeNet~\cite{codenet}, a large-scale collection of over 14 million code samples across over 50 programming languages, focusing on 4,000 problem-solving tasks from competitive programming.
Each problem \( p \) in the CodeNet dataset includes multiple submissions \( S = (s_1, s_2, ...) \) from different users and these submissions consist of both buggy and correct code snippets. Figure~\ref{example} depicts a user’s progression from an initial failed submission to a correctly revised version. 
We construct CodeNanoFix targeting examples where two submissions are textually similar but functionally different, specifically focusing on Python, the predominant language for evaluating code LLMs.
% The original data is derived from CodeNet~\cite{codenet} dataset, a large-scale collection of over 14 million code samples across more than 50 programming languages, focusing on 4000 problem-solving tasks from competitive programming. We further processed the CodeNet dataset to meet evaluation requirements, including (1)~focusing on Python programming language, (2)~verifying original labels, (3)~calculating the similarity between code samples, (4)~translating natural language descriptions from different languages to unify all descriptions. As a result, CodeNanoFix is more streamlined and consistent, making it better suited for evaluating the discriminator's ability to distinguish subtle changes in code.
% The detailed data construction process is as follows.

\subsubsection{Data Cleaning}
Data quality is crucial when building a code-related dataset. We observe several issues in the original CodeNet dataset that could potentially impact model training and evaluation. First, some code snippets in the dataset are mislabeled, such as code that fails certain test cases is incorrectly marked as functionally correct. 
To ensure label accuracy, we rerun all test cases and verify each label. 
We remove code samples where test results did not match the original labels, retaining only those where the test outputs are consistent with the original labels. Thus, we aim to maintain data consistency without introducing new labels, thereby avoiding potential label errors due to variations in test environments.
Additionally, the problem descriptions in the original dataset are in different languages, which may interfere with the model’s understanding of the task. To address this, we use ChatGPT~\cite{chatgpt2024} to translate all problem descriptions into English for uniformity. These improvements enhance the dataset’s consistency and reliability, allowing it to reflect model performance in code discrimination tasks more accurately.

\begin{table*}[t]
\centering
\renewcommand{\arraystretch}{1.1}
\caption{The statistics of the CodeNanoFix dataset.}
\label{dataset}
\scalebox{1}{
\begin{tabular}{cccccccc}
\hline
               & \#Problem & \#Sample & \begin{tabular}[c]{@{}c@{}}\#Average Tokens. \\ Problem \end{tabular} & \begin{tabular}[c]{@{}c@{}}\#Average Tokens.\\ Buggy Code\end{tabular} & \begin{tabular}[c]{@{}c@{}}\#Average Tokens.\\ Correct Code\end{tabular} & \begin{tabular}[c]{@{}c@{}}\#Average  \\ Edit Distance\end{tabular} & \begin{tabular}[c]{@{}c@{}}\#Average Relative \\ Edit Distance\end{tabular} \\ \hline
Training Set   & 480        & 80,906    & 187.64                                                               & 109.65                                                                & 112.82                                                                  & 15.52                                                               & 0.11                                                                     \\
Validation Set & 565        & 4,742     & 202.31                                                               & 198.52                                                                & 273.73                                                                  & 22.78                                                               & 0.10                                                                      \\
Test set       & 120        & 3,583     & 178.68                                                               & 129.24                                                                & 133.36                                                                  & 19.86                                                               & 0.12                                                                     \\ \hline
\end{tabular}
}
\vspace{-0.4cm}
\end{table*}

\subsubsection{Dataset Construction}
After determining the problem descriptions and the labels corresponding to individual code snippets, we construct data samples in the format \(\langle p_{i}, c^{neg}_i, c^{pos}_i \rangle\), where \(c^{neg}_i\) represents an incorrect code submission for problem \(p_{i}\), and \(c^{pos}_i\) represents a correct submission. For a complex problem \(p_{i}\), a user often needs multiple attempts to reach the correct solution. For example, with three submissions \(S = (s1, s2, s3)\), \(s3\) may be the correct code, while \(s1\) and \(s2\) contain bugs. 
To ensure that only minimal modifications are required for the model to fix buggy code, we calculate the Jaccard similarity~\cite{jaccard1901} between the correct submission \(s3\) and the buggy submissions \(s1\) and \(s2\), denoted as \(j(s1, s3)\) and \(j(s2, s3)\). We retain only those samples where the Jaccard similarity is greater than 0.9. Given the large number of users, and to ensure diversity in the constructed data samples, we group multiple users into one user group and follow the same process to create data samples.
Table~\ref{dataset} shows the basic statistics of the CodeNanoFix dataset.
The final dataset consists of training, validation, and test sets, each containing 480, 565, and 120 algorithmic problems and 80,906, 4,742, and 3,583 data samples. The validation set contains algorithmic problems that are generally more specialized, leading to a greater number of problems but fewer data samples. To avoid data leakage, we ensure that problem descriptions in the training and test sets are as distinct as possible, reducing any potential influence of training data on test data. This setup allows for effective evaluation of the model’s performance during training.

\subsubsection{Dataset Metric}
To verify data samples in the CodeNanoFix dataset requiring only minimal changes to fix bugs, we utilize the relative edit distance~(RED) metric to quantify the similarity between buggy code and correct code. We first tokenize the code using the DeepSeek~(DS)-Coder~\cite{guo2024deepseekcoder} tokenizer and calculate the number of tokens per code sample. Then, we compute the average edit distance~(ED) between buggy and correct code versions to quantify the number of modification operations required for repairs. The results indicate that the average edit distance involves significantly fewer tokens than the total token count of the code itself.
The RED metric is in the last column of the table, representing the ratio of required edit operations to the original token count~(a value between 0 and 1). The specific formula is as follows:
\[
\text{RED} = \frac{\text{ED}}{\text{AVG(B\_Tokens, C\_Tokens)}},
\]
where $AVG$ indicates the average value of two numbers, $B\_Tokens$ represents the number of tokens in the buggy code, and $C\_Tokens$ represents the number of tokens in the correct code.
The results show a typical RED of around 0.1, confirming that only minor adjustments are needed to change one code snippet to the other in the CodeNanoFix dataset.
As a result, CodeNanoFix is more streamlined and consistent, making it better suitable for evaluating the discriminator's ability to distinguish subtle changes in code.

\section{Experimental Setup}
In this section, we first present our research questions~(RQs). Then, we present the benchmark datasets, baseline models, evaluation metrics, and implementation details used to evaluate Condor. 

\subsection{Research Questions}
To evaluate the effectiveness of Condor, we ask the following questions:
\paragraph{RQ1: How well does Condor recognize subtle code differences and distinguish between textually similar but functionally different code? }
To answer this question, we apply Condor to the CodeNanoFix benchmark dataset and evaluate its performance in classification and discrimination scenarios.

\paragraph{RQ2: Can Condor generalize to other code generation tasks?} To explore this, we evaluate its performance on common code generation datasets, comparing the Pass@1 performance before and after discrimination.

\paragraph{RQ3: How do the proposed code detail capture strategies affect model performance?} To address this, we conduct ablation experiments comparing the performance of the embedding-level contrastive learning strategy and the data-level intermediate code augmentation strategy.

\subsection{Datasets}
We introduce the CodeNanoFix benchmark dataset, which contains 80,906 training samples, 4,742 validation samples, and 3,583 test samples, specifically designed to assess a model's ability to distinguish subtle code differences. This dataset focuses on code examples where only minor changes are needed for bug fixes, aiming to evaluate whether models can accurately capture code detail variations and effectively differentiate between functionally correct and incorrect code.

\lqy{
We conduct experiments on the APPS~\cite{apps}, MBPP~\cite{mbpp}, and LiveCodeBench~\cite{jain2024livecodebench} benchmark datasets to further validate Condor's generalization capabilities. The APPS dataset consists of 5,000 training problems and 5,000 test problems, covering a wide range of complex programming problems. We evaluate each baseline's original code generation accuracy and its post-process accuracy on the test set, analyzing how Condor performs in more challenging code generation tasks. The MBPP dataset contains 374 training problems, 90 validation problems, and 500 test problems, primarily used for code generation evaluation. Through experiments on the MBPP dataset, we test Condor's generalization ability and discrimination precision across problems of different scales.
In addition, we evaluate on LiveCodeBench (version 5), which comprises 880 problems focused on dynamic and real-world coding scenarios. Unlike APPS and MBPP, LiveCodeBench does not provide a dedicated training set for discriminators, making it particularly suitable for testing the generalization ability of Condor and all baselines in practical, unconstrained settings. 
}

\lqy{
By evaluating Condor on the CodeNanoFix, APPS, MBPP, and LiveCodeBench datasets, we verify the model's performance across various tasks and data scales, especially its ability to capture code details and generalize to diverse code generation tasks.
}

\subsection{Baselines}
% We categorize the baselines into two types: baselines for various discriminators and baselines for code generation. To make the discriminator more accessible, the models involved in training discriminators are typically smaller in parameter size compared to code generation models. Therefore, we select a random strategy, along with models based on CodeRanker, CodeT5, and DS-Coder-1.3B, as baselines for evaluating the trained discriminator's performance. To provide a convincing comparison for our discriminator’s performance, we use high-performing models in software engineering tasks, specifically DS-Coder-Instruct-6.7B, Meta-Llama-3.1-Instruct-8B, and Meta-Llama-3.1-Instruct-70B, as baselines for code generation.

\lqy{We categorize the baselines into two groups: those for evaluating discriminators and those for evaluating code generation performance.
% To ensure the trained discriminator is lightweight, easy to apply, and capable of fast inference, 
For the discriminator baselines, we select smaller-scale models for supervised discriminator training~\cite{codebert2020, codet52021, coderanker}, and utilize larger-scale models for few-shot discrimination. This design ensures that the trained discriminators remain lightweight, while also allowing us to assess the few-shot reasoning abilities of more powerful models.
Specifically, we include a random selection strategy, as well as models based on CodeRanker, CodeT5, and DS-Coder (1.3B, where `DS' represents `DeepSeek') as baselines to assess the trained discriminator's performance. For few-shot discrimination, we additionally consider DS-Coder-Instruct (1.3B) and Llama-3.1-Instruct (8B).
In contrast, to provide a robust upper bound and a compelling comparison for our discriminator’s effectiveness, we adopt high-performing large models in software engineering tasks as code generation baselines, including DS-Coder-Instruct (6.7B), Llama-3.1-Instruct (8B), and Llama-3.1-Instruct (70B).}

\subsubsection{Baselines for Training Discriminators}

\lqy{
For discrimination-related tasks, we adopt three main baseline settings to comprehensively evaluate model performance: (1) a heuristic-based random selection strategy, (2) discriminators trained using small-scale models, and (3) few-shot discrimination using instruction-following LLMs. 
The random strategy provides a simple baseline by randomly selecting an output from the model’s generated candidates, offering a lower bound for benchmark performance. 
For supervised discrimination, we focus on lightweight models, such as CodeRanker (built on CodeBERT), the encoder-only variant of CodeT5, and DS-Coder-1.3B. These models are selected for their proven efficiency and ease of deployment, aligning with real-world scenarios where a smaller discriminator re-ranks outputs from a larger generator. This design is particularly effective in resource-constrained environments, where deploying a large generation model alone is impractical~\cite{coderanker,codet52021,guo2024deepseekcoder}. 
CodeRanker~\cite{coderanker} fine-tunes CodeBERT~\cite{codebert2020}, an encoder model based on the BERT~\cite{bert2018} architecture, for the task of discriminating the outputs of LLMs. Trained with a masked language model on code data, CodeBERT excels in code representation learning, which enhances its applicability to various code understanding and classification tasks. 
CodeT5~\cite{codet52021} is an encoder-decoder structure that incorporates identifier-aware techniques to enhance code comprehension~\cite{divot5,twinxsql}. Following existing studies~\cite{alrashedy2024codet5_encoder, an2024codet5_encoder}, we use the encoder of CodeT5 as a baseline for training the discriminator.
DS-Coder-1.3B~\cite{guo2024deepseekcoder} is a decoder model based on the GPT architecture, trained on a large, high-quality code dataset, and performs exceptionally well in various code generation tasks, making it a solid baseline for training discriminators~\cite{grammarcoder}. 
}

\lqy{To ensure all trained baselines are compared under equally strong initial conditions, we augment the training set beyond CodeNanoFix by incorporating training data from the MBPP and APPS datasets. For these datasets, we use our code generation baselines to produce candidate outputs, which then serve as high-quality training samples for the discriminator. This strategy not only enriches the training data with diverse and challenging examples but also guarantees consistency across all discriminator baselines, thereby facilitating a fair and robust comparison.}

\lqy{
In addition, we incorporate a few-shot discrimination setting using instruction-following LLMs, specifically DS-Coder-Instruct (1.3B) and Llama-3.1-Instruct~(8B), where the model is prompted to judge code correctness or select the correct answer from multiple candidates~\cite{wang2024advanced}. The temperature is set to 0 to enforce deterministic predictions. Two common prompt formats are used: (1) ``You are given a question and a candidate answer. Your task is to determine whether the answer correctly solves or responds to the question. \textit{EXAMPLES}. \textit{TASK}.'' and (2) ``You are given a question and a list of generated outputs. Your task is to select the most likely correct answer from the list. \textit{EXAMPLES}. \textit{TASK}.'' The \textit{EXAMPLES} and \textit{TASK} denote the few-shot examples provided and the specific inference task given a particular input, respectively. This design covers both practical deployment scenarios and advanced LLM settings, enabling a thorough and fair comparison of different discrimination approaches.
}

\subsubsection{Baselines for Code Generation}
\lqy{
For code generation and repair tasks, we select high-performing, large-scale GPT-style models as baselines. DS-Coder-Instruc~(6.7B)~\cite{guo2024deepseekcoder} offers stronger code understanding and generation capabilities than its 1.3B version, making it a high-performance baseline for generation. Llama-3.1-Instruc~(8B)~\cite{dubey2024llama3}, with 8 billion parameters, is a powerful code generation model based on the LLaMA architecture~\cite{touvron2023llama,roziere2023codellama}, specifically optimized for complex code generation scenarios. Llama-3.1-Instruct~(70B), with 70 billion parameters, handles even greater code generation complexity, establishing itself as a top-performing baseline for code generation tasks.
For each input question, every model generates 10 candidate answers. In the subsequent discrimination stage, the discriminator must select the single most likely correct output from these 10 candidates as the final answer. This setup enables a direct evaluation of the discrimination model’s ability to identify the best solution among strong generative outputs.
% Notably, the upper bound of the discrimination stage corresponds to the probability that at least one of the 10 generated candidates is correct, if the discriminator could always identify the correct answer whenever it appears in the candidate set.
}

\subsection{Metrics}
To comprehensively evaluate the performance of the Condor model in code classification and discrimination tasks, we use multiple evaluation metrics, including Precision, Recall, F1-score, and Pass@1~\cite{powers2020evaluationprecisionrecallfmeasure} to measure the model’s performance before and after discriminating.

\subsubsection{Precision}
Precision measures the proportion of correctly identified functional code among all code that the model predicted as correct. It is defined as:

\[
\text{Precision} = \frac{TP}{TP + FP}
\]

where \(TP\) represents the number of correctly predicted functional code samples, and \(FP\) represents the number of incorrectly predicted functional code samples. Precision reflects the model’s performance in minimizing false positives.

\subsubsection{Recall}
Recall measures the proportion of all actual correct code that the model successfully identified. It is defined as:

\[
\text{Recall} = \frac{TP}{TP + FN}
\]

where \(FN\) represents the number of actual correct code samples that the model incorrectly classified as incorrect. Recall highlights the model’s ability to identify all correct code samples.

\subsubsection{F1-score}
The F1-score is the harmonic mean of precision and recall, combining both metrics to provide a balanced evaluation. It is defined as:

\[
\text{F1} = 2 \times \frac{\text{Precision} \times \text{Recall}}{\text{Precision} + \text{Recall}}
\]

The F1-Score is useful when precision and recall need to be balanced, offering a more comprehensive reflection of the model’s overall performance.

\subsubsection{Pass@1}
Pass@1 is an important metric in code generation tasks, representing the accuracy of the top-1 candidate code (the model's first choice). It is primarily used to assess the accuracy of code generation and is defined as:

\[
\text{Pass@1} = \frac{\text{Number of Correct Solutions on the 1st}}{\text{Total Number of Tasks}}
\]

When evaluating the model’s performance before and after discrimination, we calculate both the original Pass@1 from the generated code and the Pass@1 after applying the discrimination mechanism, comparing the performance improvement.

By combining Precision, Recall, F1-score, and Pass@1, we can comprehensively evaluate the Condor model’s performance in code classification and discrimination tasks. This approach captures the model's ability to identify correct code while also assessing improvements brought by the discrimination mechanism across different evaluation stages.

\subsection{Implementation Details}
We train two versions of Condor with different parameter scales~(i.e., Condor-110M and Condor-1.3B) based on these strategies. We choose to use CodeT5's encoder and DS-Coder~(1.3B) as the base models to train Condor at different parameter scales for two main reasons. First, both models demonstrate strong performance in various software engineering tasks and are widely used in the field, providing a solid foundation. Second, compared to larger models like DS-Coder-Instruct~(6.7B) or Llama-3.1-Instruct~(70B), they are relatively small, which helps us control the parameter size of the discriminator, making training and deployment more efficient across a variety of tasks.
% 对比学习loss中的m如何确定
The margin value $m$ for the contrastive learning loss function is set to 2 to reflect the typical magnitude of semantic differences between correct and incorrect code. 

We conduct our experiments on a machine running the Ubuntu operating system, equipped with 8 NVIDIA A100 GPUs, each with 80GB of memory, and 500GB of system memory. This hardware setup provides the necessary support for large-scale model training and dataset loading.
To ensure the reproducibility of our experiments, we implement the models and experiments using the Transformers and Huggingface~\cite{huggingface} libraries. Additionally, to accelerate the training process, we utilize DeepSpeed~\cite{deepspeed} and Accelerate~\cite{acclerate} libraries, which optimize computational efficiency in a multi-GPU environment, significantly reducing training time.

\section{Results}

\qyminor{
In this section, we represent the experimental results.
For all performance comparisons in our tables, we use the same superscript notation (e.g.,~$\alpha$) to indicate, within each column, cases where models of similar size do not show statistically significant differences from our approach.
Notably, the upper bound of the discrimination process is denoted as ``MP'' (Maximum Potential), which represents the probability that at least one of the 10 generated candidates is correct, assuming the discriminator can always identify the correct answer if it exists in the candidate set~(i.e., $\mathrm{MP} = \frac{\text{Number of tasks with at least one correct candidate}}{\text{Total number of tasks}}$). In the result tables, ``MP'' indicates the maximum achievable performance for each code generation model.
}
% Notably, the upper bound of the discrimination corresponds to the probability that at least one of the 10 generated candidates is correct, assuming the discriminator can always identify the correct answer within the candidate set. In the result tables, ``MP'' is used to denote this maximum potential for each code generation model.
% For all performance comparisons in our tables, we use the same superscript notation~(e.g., $\alpha$) to indicate cases where models of similar size do not show statistically significant differences from our approach. 

\subsection{RQ1: Effectiveness of Condor}
We evaluate Condor's effectiveness from two perspectives. First, we assess whether it can distinguish subtle code differences by treating the CodeNanoFix dataset as a classification task, determining whether the model can effectively classify between correct and incorrect code. Second, we use code LLMs to repair buggy code in the CodeNanoFix dataset and have Condor discriminate the generated outputs to see whether it enhances model performance after discrimination.
% \lqy{
% For all performance comparisons in our tables, we use the same superscript notation~(e.g., $\alpha$) to indicate cases where models of similar size do not show statistically significant differences from our approach. Notably, the upper bound of the discrimination stage corresponds to the probability that at least one of the 10 generated candidates is correct, assuming the discriminator can always identify the correct answer within the candidate set. In our tables, ``MP'' is used to denote this maximum potential for each code generation model.
% }

\subsubsection{Effectiveness on Classification Scenario}
% classification

% \begin{table}[t]
% \centering
% \renewcommand{\arraystretch}{1.1}
% \caption{Condor's Performance in Classification.}
% \label{classification}
% \begin{tabular}{cccc}
% \toprule
% Model                & Precision      & Recall         & F1-Score       \\ \midrule
% CodeRanker (110M)      & 59.42          & 77.56          & 67.29          \\
% CodeT5 (110M)        & 60.09          & 76.95          & 67.48          \\
% DeepSeek-Coder (1.3B) & 71.99          & 62.77          & 67.06          \\ \midrule
% Condor (110M)        & 65.99          & \textbf{80.04} & 72.34          \\
% Condor (1.3B)        & \textbf{74.39} & 72.40          & \textbf{73.38} \\ \bottomrule
% \end{tabular}
% \vspace{-0.5cm}
% \end{table}

\begin{table*}[t]
\centering
\renewcommand{\arraystretch}{1.1}
\caption{Condor's Performance in Classification and Discrimination Scenarios, where `FS' indicates a few-shot manner and `MP' represents the max potential. The `w/o' indicates without a specific strategy in the ablation study process.}
\label{classification_dis}
\scalebox{0.95
}{
\begin{tabular}{ccccccc}
\toprule
\multirow{2}{*}{\textbf{Model}}             & \multicolumn{3}{c}{\textbf{Classification}}               & \multicolumn{3}{c}{\textbf{Discrimination}}                                                                                                                                                                                                                           \\
                                   & Precision      & Recall         & F1-Score       & \begin{tabular}[c]{@{}c@{}}Pass@1 of DS-Coder\\ -Instruct (6.7B, MP: 75.41)\end{tabular} & \begin{tabular}[c]{@{}c@{}}Pass@1 of Llama-3.1\\ -Instruct (8B, MP: 68.66)\end{tabular} & \begin{tabular}[c]{@{}c@{}}Pass@1 of Llama-3.1\\ -Instruct (70B, MP: 83.70)\end{tabular} \\ \midrule
Original                           & 50.00              & 50.00              & 50.00              & 43.23                                                                               & 36.81                                                                             & 52.64                                                                              \\
Random                             & 50.18          & 50.93          & 50.55          & 43.23                                                                               & 36.75                                                                             & 53.39                                                                              \\
CodeRanker (110M)                  & 59.42          & 77.56          & 67.29          & 42.84                                                                               & 37.48                                                                             & 57.69                                                                              \\
CodeT5 (110M)                      & 60.09          & 76.95          & 67.48          & 42.19                                                                               & 37.65                                                                             & 56.93                                                                              \\
DS-Coder (1.3B)              & 71.99          & 62.77          & 67.06          & 43.54                                                                               & 37.32                                                                             & 58.90                                                                              \\ \midrule
DS-Coder-Instruct (1.3B, FS) & 50.88          & 25.71          & 34.16          & 43.18                                                                               & 36.70                                                                             & 53.34                                                                              \\
Llama-3.1-Instruct (8B, FS)    & 55.45          & \textbf{90.26} & 68.70          & 42.59                                                                               & 38.88                                                                             & 55.74                                                                              \\ \midrule
\textbf{Condor} (110M)                      & 65.99          & 80.04          & 72.34          & 43.73                                                                               & 39.60                                                                             & 59.64                                                                              \\
\textbf{Condor} (1.3B)                      & \textbf{74.39} & 72.40$^\alpha$          & \textbf{73.38} & \textbf{50.85}                                                                      & \textbf{45.27}                                                                    & \textbf{62.63}                                                                     \\ \midrule \midrule
\textbf{w/o} Contrastive Learning           & 73.12          & 72.06$^\alpha$          & 72.59          & 48.03                                                                               & 42.37                                                                             & 60.95                                                                              \\
\textbf{w/o} Intermediate Data              & 67.98          & 74.66          & 71.16          & 45.29                                                                               & 38.54                                                                             & 59.84                                                                              \\ \bottomrule
\end{tabular}
}
\vspace{-0.5cm}
\end{table*}

% Table~\ref{classification_dis} shows the classification performance of the models with different parameter sizes on the CodeNanoFix dataset. From the table, it can be observed that smaller models~(such as CodeRanker and CodeT5) have lower precision but higher recall, while larger models~(such as DeepSeek-Code) demonstrate higher precision but lower recall, achieving more balanced overall performance.

\lqy{
% The left side of the Table~\ref{classification_dis} summarizes the classification performance of models with varying parameter sizes on the CodeNanoFix dataset. We observe that smaller discriminative models~(such as CodeRanker and CodeT5), as well as larger models used in a few-shot setting, tend to achieve lower precision but higher recall, reflecting a tendency to over-predict positives.
The left side of the Table~\ref{classification_dis} presents the classification performance of various approaches on the CodeNanoFix dataset, including our Condor model, random selection, models trained specifically for the discrimination task, and large language models used in a few-shot setting without dedicated training. Across all evaluated metrics, Condor consistently outperforms both the random baseline and the supervised discriminators, as well as few-shot LLMs. 
}

\lqy{
For models trained specifically for discrimination, small models such as CodeRanker and CodeT5 (both 110M parameters) achieve recall rates of about 77\% but their precision is only around 60\%, resulting in F1-scores just above 67\%. This indicates that small models, due to their limited parameter capacity, tend to broadly classify code as correct in order to maximize recall. While this helps capture more true positives, it also leads to a higher rate of false positives, reducing overall precision.
In contrast, larger models like DS-Coder (1.3B) benefit from increased representational power, achieving a higher precision of about 72\% and a recall of 63\%, with a similar F1-score. The performance of larger models is relatively more balanced, as they can more precisely identify correct code while being more cautious and filtering out borderline cases.
% In comparison, Condor consistently outperforms these baselines at both scales. For example, Condor (110M) improves the F1-score to 72.34\%, and Condor (1.3B) further boosts it to 73.38\%. These results demonstrate that Condor achieves both higher precision and recall, delivering more robust discrimination performance than both CodeRanker/CodeT5 at the small scale and DeepSeek-Coder at the large scale.
}

\lqy{
Few-shot LLMs, such as Llama-3.1-Instruct (8B, FS) and DS-Coder-Instruct (1.3B, FS), show a different pattern. Without task-specific training, these models tend to favor high recall—Llama-3.1-Instruct (8B, FS) achieves a recall of over 90\%, but their precision drops to around 55\%, resulting in an F1-score below 69\%. This suggests that few-shot prompting helps these models identify most of the correct cases but also causes them to misclassify many incorrect samples as correct, leading to more false positives. DS-Coder-Instruct (1.3B, FS) performs even worse, with both precision and recall dropping below 51\% and 26\%, respectively, highlighting the limitations of relying on general generation capabilities without targeted discrimination training.
}

\lqy{
In contrast, Condor incorporates embedding-level and data-level strategies specifically designed to enhance the model’s sensitivity to code semantics and structural details. As a result, Condor (110M) substantially improves precision to 66\%, an improvement of about 6 percentage points compared to traditional discriminators like CodeRanker~(59\%) and CodeT5~(60\%). At the 1.3B scale, Condor reaches 74\% precision and 72\% recall, with an F1-score of 73\%, surpassing DS-Coder (1.3B) by more than 9 percentage points in the recall metric. These improvements are primarily due to Condor’s detail-aware representations, which allow the model to more accurately identify subtle distinctions between correct and incorrect code. This enables Condor to maintain high recall while greatly reducing false positives, thus boosting precision.
}

\subsubsection{Effectiveness on Discrimination Scenario}

\lqy{
The right side of Table~\ref{classification_dis} reports the discrimination results, measuring the functional correctness of the top-1 code selected by each discriminator using the Pass@1 metric. 
% The performance of the original generation models varies significantly, with DS-Coder Instruct 6.7B achieving a Pass@1 of 43.23\%, Llama-3.1 Instruct 8B of 36.81\%, and Llama-3.1 Instruct 70B performing the best of 52.64\%.
}

We first analyze whether subtle code differences impact model performance. Specifically, we examine label~(i.e., correct or error) consistency in Llama-3.1-Instruct~(70B) when addressing the same input with similar tokens~(Jaccard similarity bigger than 0.7). The results show that 47.55\% of such samples have inconsistent labels~(i.e., the probability of making mistakes). Separately, we investigate the Jaccard similarity for cases with label inconsistencies and find that over 37.44\% of such samples have a Jaccard similarity exceeding 0.7. 
This shows that even a 70B-parameter model can generate incorrect code due to subtle differences, underscoring the need for discriminators to capture fine-grained code details.

\lqy{
After applying discriminators to the generated outputs, we observe clear performance improvements. Notably, Condor (1.3B) consistently achieves the largest gains across all code generation models. For DS-Coder-Instruct (6.7B), Pass@1 rises from 43.23\% to 50.85\%, an increase of 17.6\% relative to the baseline. On Llama-3.1-Instruct (70B), Condor (1.3B) boosts Pass@1 from 52.64\% to 62.63\%, corresponding to a 19\% relative improvement. Condor (110M) also yields improvements across all models, though the magnitude is smaller, which reflects the importance of sufficient model capacity for capturing nuanced code features in complex generation settings.
By comparison, other discriminators such as CodeRanker and CodeT5 may offer some improvement in Pass@1, but the gains are not consistent across all settings.
The random selection baseline performs nearly identically to the original models, underscoring that naive selection strategies contribute little to overall performance.
}

% After applying a discriminator to the generated outputs, we observe that Condor~(1.3B) significantly outperforms other discriminators. 
% For example, Condor (1.3B) raises the Pass@1 score on Deepseek-Coder Instruct (6.7B) to 50.85\%, resulting in a 17.63\% increase compared to the original result.
% On Meta-Llama-3.1-Instruct~(70B), Pass@1 reaches 62.63\%, reflecting an improvement of nearly 19\% over the original result.
% While Condor (110M) also shows improvements across multiple models, the gains are less pronounced compared to their larger counterparts, highlighting the importance of parameter size in capturing complex code details. 
% In contrast, other discriminators (i.e., CodeRanker and CodeT5) also enhance the quality of generated results but yield smaller improvements, particularly in more complex generation tasks. The results of random selection show little difference from the original models, indicating that random selection has minimal impact on model performance.

These results demonstrate that Condor's incorporation of a code detail representation significantly enhances its ability to capture subtle code differences. Whether using a small or large model, Condor substantially improves the quality of selected generated code, demonstrating superior discrimination performance.

% \begin{tcolorbox}[left=0cm, right=0cm, top=0cm, bottom=0cm]
\finding{
% \textbf{Answer to RQ1}: 
The experimental results demonstrate that code details play a crucial role in improving both classification and discrimination performance. By introducing code detail-aware strategies, Condor significantly enhances its ability to capture subtle code differences, leading to improved precision and recall across different models. For example, Condor~(1.3B) achieves a precision of 74.39\% and a recall of 72.40\%. In the discrimination scenario, Condor's performance is particularly outstanding. On Llama-3.1 Instruct 70B, Condor~(1.3B) increases Pass@1 from 52.64\% to 62.63\%. This demonstrates Condor’s ability to identify correct code and exhibit strong discriminative capabilities across models of various sizes. 
}
% \end{tcolorbox}

\subsection{RQ2: Generalizability to Code Generation Tasks}
% Tables~\ref{apps} and \ref{mbpp} present the discrimination capabilities of our Condor model on the APPS and MBPP datasets, using the Pass@1 metric to measure Condor's effectiveness in selecting LLM outputs. From these experimental results, we can observe Condor's generalization ability in code generation tasks.

\lqy{
To comprehensively evaluate the generalization capability of Condor, we conduct experiments on three representative datasets: APPS, MBPP, and the recently introduced LiveCodeBench~(version 5). Table~\ref{apps}, Table~\ref{mbpp}, and Table~\ref{lcb} present the Pass@1 results of various discriminators, including both traditional baselines and our Condor model.
}

\begin{table}[t]
\centering
\renewcommand{\arraystretch}{1.1}
\caption{Condor's generalization capability in APPS, where `FS' indicates a few-shot manner and `MP' represents the max potential.}
% \vspace{-0.3cm}
\label{apps}
\scalebox{0.75}{
\begin{tabular}{cccc}
\toprule
\textbf{Model's Pass@1}              & \begin{tabular}[c]{@{}c@{}}DS-Coder-Instruct\\  (6.7B, MP: 24.58)\end{tabular} & \begin{tabular}[c]{@{}c@{}}Llama-3.1-Instruct\\  (8B, MP: 19.06)\end{tabular} & \begin{tabular}[c]{@{}c@{}}Llama-3.1-Instruct\\  (70B, MP: 35.64)\end{tabular} \\ \midrule
Original                    & 9.40                                                                   & 6.06                                                                 & 10.16                                                                 \\
Random                      &   9.60                                                                     &   6.18                                                                   &     11.74                                                                  \\
CodeRanker (110M)      & 11.56                                                                  & 9.22                                                                 & 21.30                                                                 \\
CodeT5 (110M)        & 12.18$^\alpha$                                                                  & 9.36                                                                 & 21.62                                                                 \\
DS-Coder (1.3B) & 12.10                                                                  & 9.64                                                                 & 22.42                                                                 \\ \midrule
DS-Coder-Instruct (1.3B, FS) & 9.42                                                                     & 6.00                                                                   & 10.12                                                                   \\
Llama-3.1-Instruct (8B, FS)    & 10.18                                                                    & 6.40                                                                   & 9.80                                                                    \\ \midrule

\textbf{Condor} (110M)               & 12.24$^\alpha$                                                                  & 10.32                                                                & 22.32                                                                 \\
\textbf{Condor} (1.3B)               & \textbf{14.68}                                                         & \textbf{12.08}                                                       & \textbf{25.10}                                                        \\ \bottomrule
\end{tabular}
}
\vspace{-0.2cm}
\end{table}

\begin{table}[t]
\centering
\renewcommand{\arraystretch}{1.1}
\caption{Condor's generalization capability in MBPP, where `FS' indicates a few-shot manner and `MP' represents the max potential.}
% \vspace{-0.3cm}
\label{mbpp}
\scalebox{0.75}{
\begin{tabular}{cccc}
\toprule
\textbf{Model's Pass@1}              & \begin{tabular}[c]{@{}c@{}}DS-Coder-Instruct\\ (6.7B, MP: 79.2)\end{tabular} & \begin{tabular}[c]{@{}c@{}}Llama-3.1-Instruct\\ (8B, MP: 77.2)\end{tabular} & \begin{tabular}[c]{@{}c@{}}Llama-3.1-Instruct\\ (70B, MP: 86.2)\end{tabular} \\ \midrule
Original                    & 59.00                                                                  & 54.00                                                                & 71.40                                                                 \\
Random                      &   59.60                                                                     &  55.20                                                                &             71.60                                                              \\
CodeRanker (110M)       & 63.20$^\alpha$                                                                  & 56.60                                                                & 72.20                                                                 \\
CodeT5 (110M)         & 61.60                                                                  & 58.20                                                                & 72.80$^\beta$                                                                 \\
DS-Coder (1.3B)  & 64.20                                                                  & 57.60                                                                & 73.60                                                                 \\ \midrule

DS-Coder-Instruct (1.3B, FS) & 59.20                                                                    & 54.20                                                                  & 71.40                                                                   \\
Llama-3.1-Instruct (8B, FS)    & 59.60                                                                    & 55.00                                                                  & 71.20                                                                   \\ \midrule

\textbf{Condor} (110M)               & 63.20$^\alpha$                                                                  & \textbf{60.40 }                                                               & 73.00$^\beta$                                                                 \\
\textbf{Condor} (1.3B)               & \textbf{65.00}                                                         & 59.00                                                       & \textbf{75.20}                                                        \\ \bottomrule
\end{tabular}
}
% \vspace{-0.3cm}
\end{table}

\begin{table}[t]
\centering
\renewcommand{\arraystretch}{1.1}
\caption{Condor's generalization capability in LiveCodeBench, where `FS' indicates a few-shot manner and `MP' represents the max potential.}
% \vspace{-0.3cm}
\label{lcb}
\scalebox{0.75}{
\begin{tabular}{cccc}
\toprule
\textbf{Model's Pass@1}                     & \begin{tabular}[c]{@{}c@{}}DS-Coder-Instruct\\  (6.7B, MP: 18.86)\end{tabular} & \begin{tabular}[c]{@{}c@{}}Llama-3.1-Instruct\\  (8B, MP: 15.11)\end{tabular} & \begin{tabular}[c]{@{}c@{}}Llama-3.1-Instruct\\  (70B, MP: 35.11)\end{tabular} \\ \midrule
Original                           & 15.68                                                                    & 12.73                                                                  & 28.97                                                                   \\
Random                             & 15.22                                                                    & 12.15                                                                  & 28.18                                                                   \\
CodeRanker (110M)                  & 15.11                                                                    & 12.50                                                                  & 29.09                                                                   \\
CodeT5 (110M)                      & 15.34$^\alpha$                                                                    & 12.50                                                                  & 29.43                                                                   \\
DS-Coder (1.3B)              & 15.40                                                                    & 12.84$^\beta$                                                                  & 29.68                                                                   \\ \midrule
DS-Coder-Instruct (1.3B, FS) & 15.80                                                                    & 12.72                                                                  & 28.86                                                                   \\
Llama-3.1-Instruct (8B, FS)    & 15.68                                                                    & 12.05                                                                  & 29.55                                                                   \\ \midrule
\textbf{Condor} (110M)                      & 15.68$^\alpha$                                                                    & 13.07$^\beta$                                                                  & 30.45                                                                   \\
\textbf{Condor} (1.3B)                      & \textbf{16.25}                                                           & \textbf{13.18$^\beta$}                                                         & \textbf{30.90}                                                          \\ \bottomrule
\end{tabular}
}
% \vspace{-0.2cm}
\end{table}

In the APPS dataset~(Table~\ref{apps}), due to the challenging nature of the algorithmic problems, the original generation models show relatively low performance. DS-Coder Instruct 6.7B achieves a Pass@1 of 9.40\%, Llama-3.1 Instruct 8B reaches 6.06\%, and the Llama-3.1 Instruct 70B version achieves 10.16\%. The results of random selection differ little from the original outputs, for example, on DS-Coder-Instruct 6.7B, Pass@1 only increases from 9.40\% to 9.60\%. In contrast, traditional discriminators like CodeRanker (110M) and CodeT5 (110M) show more significant improvements, with CodeRanker (110M) boosting Llama-3.1-Instruct 70B's Pass@1 to 21.30\%, and CodeT5 (110M) slightly higher at 21.62\%. As the model size increases, DS-Coder~(1.3B) further improves Pass@1 to 22.42\%, demonstrating solid discriminative ability.
Condor, enhanced by its code detail-awareness strategies, performs exceptionally well on the APPS dataset. Condor~(110M) surpasses CodeRanker and CodeT5, increasing Llama-3.1 Instruct 8B's Pass@1 to 10.32\%, showcasing its discrimination advantage even in smaller models. Condor (1.3B) performs the best, achieving the highest Pass@1 across all models. For example, on the Llama-3.1-Instruct~(70B), it boosts Pass@1 to 25.10\%, representing an improvement over 147\% from the original 10.16\%. This demonstrates Condor's strong ability to capture code details and significantly enhance performance in discrimination tasks.

\begin{figure}[t]
  \centering
  % \vspace{-0.13cm}
  % \setlength{\abovecaptionskip}{10pt}
  \includegraphics[scale=0.75]{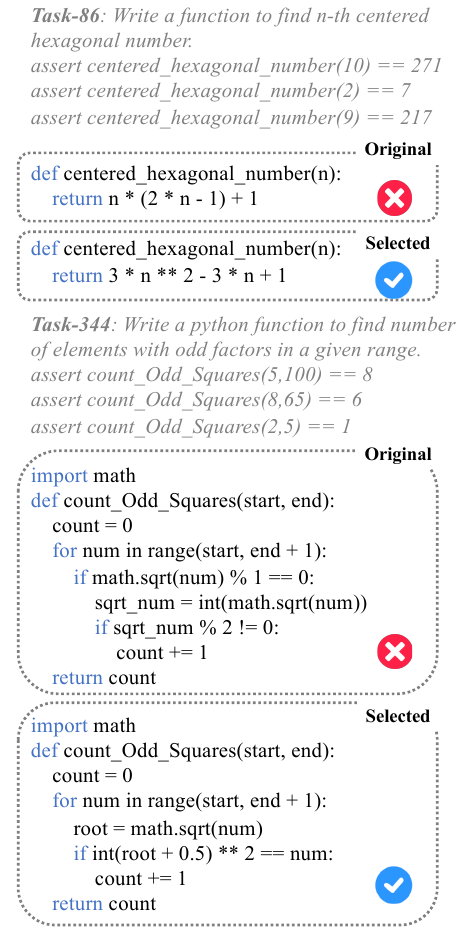}
  \caption{Case studies showing Condor’s ability to recognize correct solutions and detect errors in code.}
  \label{cases}
  % \vspace{-0.2cm}
\end{figure}

The experiments on the MBPP dataset~(Table~\ref{mbpp}) further validate Condor's generalization ability. Due to the relatively simpler nature of the MBPP problems, the original generation models' Pass@1 scores are higher across the board. For example, DS-Coder Instruct 6.7B achieves a Pass@1 of 59.00\%, Llama-3.1-Instruct 8B reaches 54.00\%, and Llama-3.1-Instruct 70B reaches 71.40\%. Similar to the APPS dataset, random selection yields almost no improvement. CodeRanker (110M) and CodeT5 (110M) show more notable gains in discrimination performance, with CodeT5 (110M) improving Llama-3.1 Instruct 70B's Pass@1 to 72.80\%, slightly higher than CodeRanker (110M) at 72.20\%. DS-Coder (1.3B) performs steadily on this dataset, raising Pass@1 to 73.60\%.
Similarly, Condor performs outstandingly on the MBPP dataset, especially Condor (1.3B), which further improves Llama-3.1 Instruct 70B's Pass@1 to 75.20\%, the best result. Condor (110M) also performs well on smaller models, boosting Llama-3.1 Instruct 8B's Pass@1 to 60.40\%, surpassing DS-Coder and other discriminators. Overall, Condor demonstrates competitiveness not only in smaller models but also in larger models, well utilizing its ability to capture code details and significantly improving Pass@1 accuracy.

\lqy{
Figure~\ref{cases} presents two cases in MBPP dataset, highlighting Condor’s discrimination abilities at different granularity levels. 
In the first example~(top, Task-86), Condor distinguishes between two formulas for the centered hexagonal number and correctly identifies the accurate computation based on fine-grained numerical differences. 
In the lower example~(bottom, Task-344), the task requires counting the number of elements with odd factors in a given range. Condor is faced with a more challenging scenario involving logic-level reasoning. The first candidate solution checks for perfect squares and then further filters those whose square roots are odd, which does not align with the problem’s requirement of counting all perfect squares (since every perfect square has an odd number of factors, regardless of the parity of the square root). Condor successfully recognizes the discrepancy in the logical process and selects the correct solution, showing its capability to capture not only numerical but also coarse-grained semantic differences. These cases highlight that Condor can accurately discriminate between candidates based on both precise computational correctness and broader algorithmic logic, supporting its robust performance across a variety of code reasoning scenarios.
}

\lqy{
Following our analysis of the APPS and MBPP, we further evaluate Condor on LiveCodeBench~(Table~\ref{lcb}), a challenging new benchmark that emphasizes dynamic and real-world code tasks. Importantly, LiveCodeBench does not provide a dedicated training set for discriminators, making it more challenging for evaluating true model generalization and real-world applicability.
The original Pass@1 scores are moderate (e.g., DS-Coder-Instruct 6.7B: 15.68\%, Llama-3.1-Instruct 8B: 12.73\%, Llama-3.1-Instruct 70B: 28.97\%), with random selection again yielding almost no gain. The improvements brought by traditional discriminators are modest, and their effects are even less pronounced for this more complex dataset.
Instead, Condor continues to show a strong generalization advantage. For example, Condor (1.3B) pushes Llama-3.1-Instruct 70B's Pass@1 to 30.90\%, the highest among all tested discriminators, while smaller-scale Condor models also surpass the other baselines on various settings. 
Although the gains on LiveCodeBench are smaller than on APPS, this is because the maximum potential (MP) for improvement is lower, as shown in these tables~(i.e., `MP' in Table \ref{apps},~\ref{mbpp},~\ref{lcb}). With fewer correct candidates generated, even the best discriminator can provide only limited gains, highlighting the challenge of this benchmark.
}

% \begin{tcolorbox}[left=0cm, right=0cm, top=0cm, bottom=0cm]
% \textbf{Answer to RQ2}:
\finding{
The experimental results demonstrate that Condor exhibits outstanding generalization capabilities, effectively improving discrimination accuracy on the APPS, MBPP, and LiveCodeBench datasets. On APPS, Condor (1.3B) raises Llama-3.1-Instruct 70B's Pass@1 from 10.16\% to 25.10\%. On MBPP, Condor (1.3B) boosts Pass@1 from 71.40\% to 75.20\%. Similarly, on LiveCodeBench, Condor consistently delivers the highest Pass@1 among all tested discriminators, further validating its robustness and effectiveness across diverse and challenging benchmarks.
}
% \end{tcolorbox}

\subsection{RQ3: Ablation Study}

\lqy{
The lower part of Table~\ref{classification_dis} presents the ablation study results for our Condor model, comparing the full model to its variants with individual strategies removed. We use the 1.3B version of Condor for ablation because it achieves the most balanced and robust performance across all metrics, with a precision of 74.39\%, recall of 72.40\%, and F1-score of 73.38\%. In discrimination tasks, Condor (1.3B) also achieves a leading Pass@1 of 62.63\% on Llama-3.1-Instruct (70B), underscoring its effectiveness in capturing code details and serving as a strong baseline for analyzing the impact of each strategy.
}

When the embedding-level contrastive learning strategy is removed, precision drops to 73.12\%, and recall also decreases to 72.06\%, resulting in a lower F1-score of 72.59\%. This indicates that contrastive learning plays a crucial role in enhancing the model’s ability to capture functional code differences. The discrimination performance also declines, with Pass@1 on Llama-3.1-Instruct (70B) dropping from 62.63\% to 60.95\%, demonstrating the significant contribution of contrastive learning to discrimination tasks.
When the data-level intermediate data augmentation strategy is removed, the model shows more pronounced changes. Although recall increases slightly to 74.66\%, suggesting the model becomes more conservative in recognizing correct code, precision drops significantly to 67.98\%, leading to an overall F1-score decrease to 71.16\%. The discriminative performance also declines, with Pass@1 on Llama-3.1-Instruct (70B) falling to 59.84\%. This demonstrates the critical role of the intermediate data strategy in improving the model's overall performance.

\begin{figure*}[t]
  \centering
  % \vspace{-0.13cm}
  % \setlength{\abovecaptionskip}{10pt}
  \includegraphics[scale=0.7]{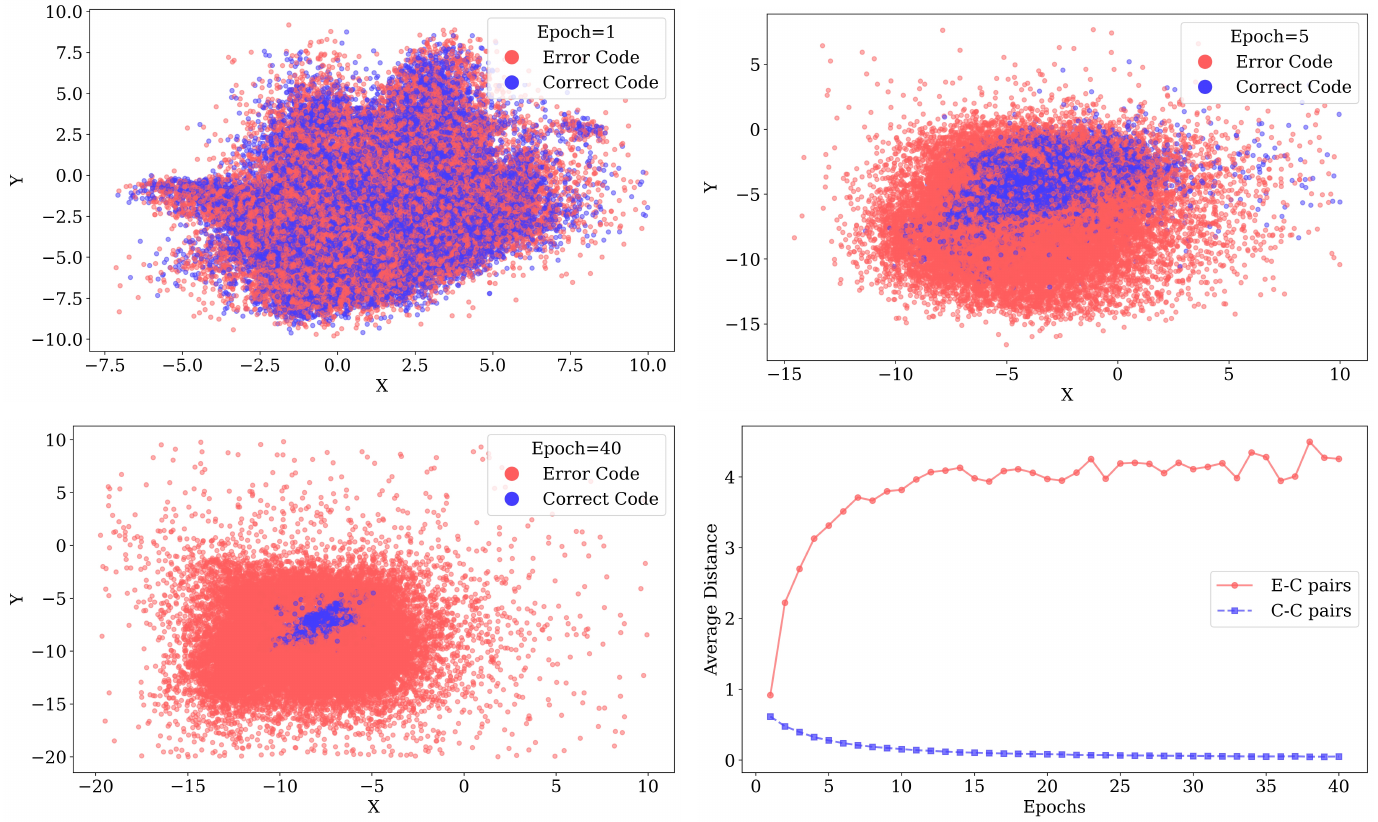}
  \caption{Illustration of the impact of contrastive learning on code representations. The first three subplots show the 2D embeddings of code after training for 1 epoch (top left), 5 epochs (top right), and 40 epochs (bottom left). The bottom right subplot illustrates the changes in the average distance between correct code snippets and between error and correct code snippets as the number of training epochs increases.}
  \label{contrastive}
  % \vspace{-0.5cm}
\end{figure*}

Figure~\ref{contrastive} visually demonstrates the impact of contrastive learning on model embedding representations. The subplots in the top left, top right, and bottom left show the progression of code embeddings projected in 2D space as training epochs increase. 
These projections, recorded before calculating the loss during contrastive learning, reveal how the embeddings evolve and separate as the model learns to distinguish correct from incorrect code.
From these subplots, we observe that as training deepens, the embeddings of correct code snippets gradually cluster closer to the center, while the embeddings of buggy code samples become more dispersed, moving farther from the center of the correct code. This indicates that contrastive learning effectively brings representations of correct code closer together while increasing the distance between correct and error code, enhancing the model’s ability to discriminate code correctness.
The bottom right subplot shows the average distances between correct code pairs~(C-C) and between error and correct code pairs~(E-C). As training progresses, the C-C distance gradually decreases, while the E-C distance increases. This trend further demonstrates that the contrastive learning strategy effectively assists the model in distinguishing fine-grained differences between correct and incorrect code representations at the embedding level.

Leveraging both contrastive learning and intermediate data augmentation strategies to capture the code details, the full Condor~(1.3B) model achieves a balance between high precision and recall in classification tasks and delivers optimal performance in discrimination tasks. When these strategies are removed, the model's ability to capture code details and Pass@1 accuracy declines, confirming the positive impact of both strategies on Condor's final performance.

% \begin{tcolorbox}[left=0cm, right=0cm, top=0cm, bottom=0cm]
% \textbf{Answer to RQ3}:
\finding{
The ablation study results show that both the contrastive learning and intermediate data augmentation strategies effectively help the model capture detailed information in the code, significantly enhancing overall model performance.
When contrastive learning is removed, both the model’s F1-score and its discrimination performance decline, indicating that contrastive learning is crucial for capturing functional code differences. 
Similarly, removing the intermediate data augmentation strategy slightly improves recall but significantly reduces precision by 8.6\% and overall discrimination performance, highlighting this strategy's crucial role in enhancing the model's accuracy and discriminative ability.
}
% \end{tcolorbox}

\section{Threats to Validity}
\textbf{Threats to internal validity} might come from the implementation of our experiments. To mitigate this threat, we use open-source models available on HuggingFace~\cite{huggingface} to minimize potential implementation errors. Additionally, to ensure the reproducibility of our experiments, we fix the random seed, ensuring consistency across different experimental results and reducing interference from randomness.

\textbf{Threats to external validity} may arise from fourth factors. The first factor may come from the choice of the base models. To mitigate this risk, we select the widely used and representative models CodeT5 and DS-Coder. CodeT5 is recognized as a classic encoder-decoder framework that has been validated and applied across multiple tasks, demonstrating strong versatility. DS-Coder is a decoder model that performs exceptionally well in a variety of software engineering tasks, providing a solid and reliable foundation for our experiments with our Condor model. 
\qyminor{
The second potential threat may arise from the choice of downstream datasets in our experiments. To mitigate this risk, we select the widely used code generation datasets, such as APPS, MBPP, and LiveCodeBench. These datasets are extensively used to evaluate code generation tasks and are commonly used to assess the performance of LLMs, providing strong support for the robustness of our experimental results. In the future, we plan to extend our evaluation to datasets beyond code generation and code repair tasks, such as ChangeGuard~\cite{groninger2025changeguard}, which focuses on functional change detection in code edits.
}
\lqy{
The third potential threat may come from the choice of programming language. Our current experiments are conducted exclusively on Python, which, while widely used in AI research and competitive programming, does not fully represent the diversity of programming languages used in real-world software engineering (e.g., Java, C++, JavaScript, Go, Rust). To mitigate this, we design Condor’s architecture to be language-agnostic, allowing it to be adapted to other languages by collecting appropriate training data and re-tokenizing. Exploring multi-language settings remains an important direction for future work.
The fourth potential threat might arise from the gap between benchmark settings and real-world development environments. In large-scale software engineering, correctness often depends on the broader system context, including dependencies and cross-module interactions, which are not captured when evaluating isolated code snippets. To mitigate this gap, we introduce LiveCodeBench into our evaluation. LiveCodeBench contains real-world, dynamic coding tasks without a public training set, making it a strict test of model generalization beyond the training distribution. Moreover, our results show that Condor scales well—larger versions (e.g., 1.3B parameters) achieve more promising performance gains, indicating potential for further improvements when moving towards more complex, production-scale scenarios.
}

\textbf{Threat to construct validity} may arise from the choice of evaluation metrics. To mitigate this, we use widely accepted metrics: precision, recall, and F1-score for classification tasks to comprehensively assess the model's discriminative ability. For discrimination tasks, we apply the Pass@1 metric, commonly used in the code generation domain, to evaluate the model's post-selection performance. These metrics effectively measure the model’s performance across different tasks, ensuring comprehensive and reliable evaluations.

\section{Related Work}

\subsection{Code Discriminators}
Existing approaches that improve code generation performance by selecting the outputs of LLMs can be categorized into two types: execution-based approaches and non-execution-based approaches. 
Execution-based approaches require executing the code and then using the feedback from the execution as additional information to enhance the discriminative effectiveness.
Among them, MBR-EXEC, CodeT~\cite{codet,mbrexec} requires executing the code to select the LLM's outputs. 
However, in real-world development scenarios, it is not always easy to obtain the necessary test cases or testing environments for execution. Moreover, executing code generated by LLMs often raises safety concerns. Consequently, execution-based approaches face significant challenges in terms of flexibility.
For flexibility, we focus on the non-execution-based approaches. 
% Coder-Reviewer reranking~\cite{coder_reviewer_reranking} uses prompt engineering by feeding the generated code to the original model to generate descriptions. It then selects the code by calculating the similarity between the generated description and the original description. 
% However, as the current code model is quite large, 
% making it costly to use the original model to assist in selecting generated outputs, which falls short of the efficiency requirements for real-world code generation tasks. Additionally, this approach is highly sensitive to prompt design and the model’s ability to interpret natural language, especially when problem descriptions are complex~(e.g., APPS dataset). This discrepancy in the complexity of generating problem descriptions versus code can significantly impact the accuracy and consistency of the generated results.
\qyminor{Coder-Reviewer reranking~\cite{coder_reviewer_reranking} employs prompt engineering by feeding the generated code back into the original model to generate descriptions, and then selects candidate code based on the similarity between the generated and original descriptions. However, as current code models are quite large, using the original model for reranking is computationally expensive and fails to meet the efficiency requirements of real-world code generation tasks. Furthermore, this approach is highly sensitive to prompt design and the model’s ability to interpret natural language, especially when problem descriptions are complex~(e.g., APPS dataset). The gap between the complexity of generating problem descriptions and generating code can lead to significant variability in the accuracy and consistency of the reranking results.
CodeRanker~\cite{coderanker}, which is similar to our work, improves model accuracy by incorporating a smaller classification model. Specifically, CodeRanker fine-tunes the CodeBERT model for code classification tasks, which in turn enhances the discrimination of the model's outputs.}
% Unlike existing models that simply leverage the inherent capabilities of models to represent code, Condor employs contrastive learning and data augmentation techniques to better detect subtle code details, thereby enhancing the discrimination performance.
\qyminor{Unlike these approaches, Condor adopts a code-centric perspective. Instead of relying on natural language interpretation or simple code representations, Condor leverages contrastive learning and data augmentation to directly analyze the structural and semantic differences between code candidates. This enables Condor to better detect subtle variations in code details that may significantly impact program functionality, thereby providing more robust and efficient discrimination of correct solutions.}

\subsection{Datasets for Code Discrimination}
Current code discriminators commonly use code generation datasets as evaluation benchmarks. For example, CodeRanker, CodeT~\cite{coderanker, codet} have been evaluated on APPS, MBPP, HumanEval~(i.e., code generation datasets for Python programming language)~\cite{apps, mbpp, humaneval}, and MBR-EXEC, Coder-Reviewer reranking~\cite{mbrexec,coder_reviewer_reranking} have been evaluated on Spider~(i.e., generation dataset for SQL code) and NL2Bash~(i.e., generation dataset for Bash code )~\cite{spider,nl2bash}.
% ChangeGuard~\cite{groninger2025changeguard} have been evaluated to assess whether code changes in real-world edits actually alter program functionality. 
% When applied to such datasets, the model may generate candidate samples for the same problem that vary significantly at the textual level, because a correct code can be expressed in various textual forms.
Although these datasets can be used to evaluate the overall performance of discriminators, they fall short in assessing the discriminators's true potential to detect subtle code details. 
In code generation datasets, inputs are typically natural language descriptions prompting the model to generate corresponding code. For the same functionality, there may be multiple implementation approaches, so the generated code is not always textually similar. Additionally, these datasets rely on outputs generated by LLMs, making their effectiveness dependent on the specific LLM used. 
Therefore, we construct the CodeNanoFix dataset from real user-submitted code, focusing on detailed changes in the code repair process to provide a more direct and effective evaluation of the discriminator's performance in distinguishing code details.
% In contrast, the code to be selected in CodeNanoFix typically contains only subtle differences, which better reflects the model's discriminative ability to handle code details. 

\subsection{Code Repair Datasets}
There are some studies that focus on the construction of code repair datasets, such as Defects4J, Bugs.jar, Bugs2Fix, QuixBug, IntroClass, HumanEvalPack~\cite{defects4j, bugsjar, codexglue, quixbugs, introclass, commitpack_muennighoff2023octopack}. 
However, these datasets are not designed to evaluate the discriminator's ability to detect subtle code details. The reasons are as follows. 
(1) For repository-level repair datasets like Defects4J and Bugs.jar~\cite{defects4j,bugsjar}, the process involves multiple stages, including defect localization, code repair, and patch detection. LLMs can hardly directly generate a series of code candidates to discriminate without specific fine-tuning or prompt design. 
% Moreover, the modified code may involve extensive change operations, rather than the subtle differences that are the focus of our dataset. 
In contrast, our CodeNanoFix dataset includes independent problem descriptions and code, allowing LLMs to directly generate fixed code for buggy code. Additionally, CodeNanoFix focuses on fine-grained code differences rather than extensive modifications, making it easier to evaluate the model’s potential.
(2) The Bugs2Fix, QuixBug, IntroClass, and HumanEvalPack datasets are function-level repair datasets, but they are still not suitable for evaluating the discriminators. The Bugs2Fix~\cite{codexglue} is a Java code repair dataset that includes erroneous Java code with variable anonymization. Due to the anonymization and lack of natural language descriptions of the code's functionality, this dataset focuses more on syntactic and type errors rather than semantic errors. 
The QuixBug dataset~\cite{quixbugs} contains 40 erroneous Python code samples, with only a test set and no training set, aimed at evaluating code repair effectiveness. 
However, this dataset focuses on different repair problems and lacks fine-grained variations of code for the same repair task, making it challenging to evaluate the capabilities of discriminators.
The IntroClass dataset~\cite{introclass} provides tens of thousands of erroneous C language answers for six programming tasks, while CodeNanoFix is sourced from thousands of different problems, offering greater semantic diversity.
The HumanEvalPack~\cite{commitpack_muennighoff2023octopack} dataset is based on 164 problems from the HumanEval~\cite{humaneval} dataset, with each problem containing a manually injected bug. 
However, these samples are created by intentionally introducing errors into correct code, rather than reflecting real-world bug-fixing scenarios, making it difficult to capture the true nature of code repair. 
Additionally, each problem in this dataset corresponds to only one bug and its corrected version, lacking coverage of diverse repair contexts for the same problem and thus failing to effectively showcase the variety of potential fixes.
Therefore, we introduce a novel repair dataset, CodeNanoFix, that addresses the limitations of existing datasets in evaluating a model's ability to recognize subtle code differences.
In the CodeNanoFix dataset, the differences between the buggy and fixed code are minimal, enabling a more effective assessment of the model's discriminative capability.

\section{Conclusion}
We propose Condor, a code discriminator that focuses on fine-grained code differences. To enhance the model's ability to discriminate subtle changes, we propose an embedding-level code detail contrastive strategy and a data-level code detail augmentation strategy.
Specifically, we first employ a contrastive learning approach to optimize the code representations of the original base model, making it more sensitive to changes in code details. Then, we extract intermediate modification data from the code revision process to further enrich the training data for the discriminator, enhancing its ability to discern code details.
To evaluate the performance of the discriminator, we construct the CodeNanoFix dataset, which consists of buggy code resulting from minor differences and their corresponding correct versions. This dataset is used to assess the capabilities of various discriminators. Experimental results demonstrate that Condor outperforms other discriminators in classification performance and excels at selecting generated outputs from LLMs. Additionally, Condor is applied to discrimination tasks on other code-generation datasets, with experimental results demonstrating strong generalization capabilities, as indicated by a significantly higher Pass@1 metric compared to other models. Ablation studies further confirm that both strategies employed by Condor positively impact its overall performance.

\section{Acknowledgment}
This work is sponsored by the National Key Research and Development Program of China under Grant No. 2023YFB4503803, the National Natural Science Foundation of China under Grant No. 62232003, No. 62402482, and No. W2411051.

% This work is sponsored by the National Key Research and Development Program of China under Grant No. 2023YFB4503803, the National Natural Science Foundation of China under Grant No. 62232003

% \section{Data Availability}
% Our code and dataset are available at the link: \url{https://github.com/LIANGQINGYUAN/Condor}

\bibliographystyle{IEEEtran}
\bibliography{ref.bib}

\end{document}